\documentclass[a4paper,10pt]{article}
\usepackage{blois,epsfig}

\def\be{\begin{equation}}
\def\ee{\end{equation}}
\def\bea{\begin{eqnarray}}
\def\eea{\end{eqnarray}}

\usepackage{amsmath,amsfonts}
\newcommand{\ud}{\mathrm{d}}
\newcommand{\ve}{\varepsilon}

\begin{document}
\vspace*{4cm}
\title{SCALAR FIELD COSMOLOGY --- TOWARD DESCRIPTION OF \\ 
DYNAMIC COMPLEXITY OF COSMOLOGICAL EVOLUTION}

\author{OREST HRYCYNA$^{a}$, MAREK SZYD{\L}OWSKI$^{b,d}$ and ADAM
KRAWIEC$^{c,d}$}
\address{$^{a}$Department of Theoretical Physics,
The John Paul II Catholic University of Lublin, \\
Al. Rac{\l}awickie 14, 20-950 Lublin, Poland \\
$^{b}$Astronomical Observatory, Jagiellonian University, Orla 171, 30-244
Krak{\'o}w, Poland \\
$^{c}$Institute of Economics and Management, Jagiellonian University, \\
{\L}ojasiewicza 4, 30-348 Krak{\'o}w, Poland \\
$^{d}$Mark Kac Complex Systems Research Centre, Jagiellonian University, \\
Reymonta 4, 30-059 Krak{\'o}w, Poland}

\maketitle\abstracts{
We study the dynamical evolution of cosmological models with the 
Robertson-Walker symmetry with a scalar field non-minimally coupled to 
gravity and barotropic matter. For this aim we use dynamical system methods. 
We have found a type of evolutional path which links between all important 
events during the evolution, the cosmological singularity of finite time, 
inflation, radiation and matter dominating epoch and the accelerated phase 
expansion of the universe. We point out importance of finding the new generic 
solution called a twister solution for a deeper description of the evolution 
of the Universe. We demonstrate that including the non-minimal coupling
leads to a new, richer evolutional cosmological scenario in comparison to the 
case of minimal coupling.}

\section{Introduction}

The standard method of description matter content in cosmology bases on the 
concept of perfect fluid approximation, where pressure and energy density 
satisfy the equation of state. Different cosmological epochs constitute 
solutions of the Einstein equations corresponding different forms of the 
equation of state, usually postulated in a linear form with respect to the 
energy density. On the other hand if we study very early stages evolution 
of the Universe then characterization of matter content in terms of the 
barotropic equation of state is not adequate. In the quantum epoch including 
matter in the form of scalar field with the potential seems to be more 
suitable. We propose to describe matter in the form of both barotropic 
matter and single scalar field with the potential. We also assumed that 
there is present a nonzero coupling constant scalar field to the gravity.

A non-minimal coupling appeared naturally in quantum theory of the scalar 
field as generated by quantum corrections or required by renormalization 
of the theory.\cite{Faraoni:1996rf} The value of this coupling constant 
becomes important for cosmology. Recently, it has been constructed an 
extended model of inflation with a non-minimal coupling between the inflaton 
field and the Ricci scalar curvature.\cite{Bezrukov:2007ep,DeSimone:2008ei,Barvinsky:2008ia,Pallis:2010wt}
It was shown in particular that the non-minimal inflation can be realized 
within the Standard Model (SM) or minimal extension of it.\cite{Lerner:2009xg} 
In the SM the Higgs boson inflation and dark matter is considered by Clark et al.\cite{Clark:2009dc}
The non-minimal coupling plays also an important role in the context of 
description of quintessence epoch.\cite{Szydlowski:2008in,Hrycyna:2009zf,Hrycyna:2010yv}

The main aim of this paper is to present a generic solution of the non-minimal 
coupling cosmology called the twister solution because of the shape of its
trajectory in the phase space. We explore here methods of dynamical systems 
because their advantage of representing all solutions for all admissible 
initial conditions. The phase space is organized by critical points which 
represent stationary solutions for which right-hand sides of the dynamical 
system vanish and trajectories joining them represent the evolution of the 
system. A new type of evolution appears only if the coupling constant is 
different from minimal and conformal. This solution leads naturally to the 
quintessence epoch through the twister solution. We characterize properties 
of this acceleration by calculation of so called state-finder parameters.

We assume the spatially flat FRW universe filled with the non-minimally 
coupled scalar field and barotropic fluid with the equation of the state 
coefficient $w_{m}$. The action is
\begin{equation}
S = \frac{1}{2}\int \ud^{4}x \sqrt{-g} \left[\frac{1}{\kappa^{2}}R - \ve
\Big(g^{\mu\nu}\partial_{\mu}\phi\partial_{\nu}\phi + \xi R \phi^{2}\Big) -
2U(\phi) \right] + S_{m},
\end{equation}
where $\kappa^{2}=8\pi G$, $\ve = +1,-1$ corresponds to canonical and phantom
scalar fields, respectively, the metric signature is $(-,+,+,+)$,
$R=6\left(\frac{\ddot{a}}{a}+\frac{\dot{a}^{2}}{a^{2}}\right)$ is the Ricci scalar, $a$ is
the scale factor and a dot denotes differentiation with respect to the
cosmological time $t$ and $U(\phi)$ is the scalar field potential function. $S_{m}$
is the action for the barotropic matter part.

The dynamical equation for the scalar field we can obtain from the variation
$\delta S/\delta \phi = 0$
\begin{equation}
\ddot{\phi} + 3 H \dot{\phi} + \xi R \phi + \ve U'(\phi) =0,
\end{equation}
and energy conservation condition from the variation $\delta S/\delta g^{\mu\nu}=0$
\begin{equation}
\mathcal{E}= \ve \frac{1}{2}\dot{\phi}^{2} + \ve3\xi H^{2}\phi^{2} + \ve3\xi H
(\phi^{2})\dot{} + U(\phi) + \rho_{m} - \frac{3}{\kappa^{2}}H^{2}.
\end{equation}
Then the Einstein equation for the flat FRW model and the conservation condition read
\begin{equation}
\frac{3}{\kappa^{2}}H^{2} = \rho_{\phi} + \rho_{m}, \quad \textrm{and} \quad
\dot{H} = -\frac{\kappa^{2}}{2}\Big[(\rho_{\phi}+p_{\phi}) +
\rho_{m}(1+w_{m})\Big]
\end{equation}
where the energy density and the pressure of the scalar field are
\begin{eqnarray}
\rho_{\phi} & = & \ve\frac{1}{2}\dot{\phi}^{2}+U(\phi)+\ve3\xi H^{2}\phi^{2} +
\ve 3\xi H (\phi^{2})\dot{},\\
p_{\phi} & = & \ve\frac{1}{2}(1-4\xi)\dot{\phi}^{2} - U(\phi) + \ve\xi
H(\phi^{2})\dot{} - \ve2\xi(1-6\xi)\dot{H}\phi^{2} - 
\ve3\xi(1-8\xi)H^{2}\phi^{2} + 2\xi\phi U'(\phi).
\end{eqnarray}

\section{Non-minimal scalar field cosmology as a dynamical system}

In what follows we introduce the energy phase space variables
$x\equiv \frac{\kappa \dot{\phi}}{\sqrt{6}H}$,
$y\equiv\frac{\kappa\sqrt{U(\phi)}}{\sqrt{3}H}$,
$z\equiv\frac{\kappa}{\sqrt{6}}\phi$,
which are suggested by the conservation condition
\begin{equation}
\frac{\kappa^{2}}{3H^{2}}\rho_{\phi} + \frac{\kappa^{2}}{3H^{2}}\rho_{m} =
\Omega_{\phi} + \Omega_{m} = 1.
\end{equation}

The acceleration equation can be rewritten to the form
\begin{equation}
\dot{H} = -\frac{\kappa^{2}}{2}\Big(\rho_{\rm{eff}}+p_{\rm{eff}}\Big) =
-\frac{3}{2}H^{2}(1+w_{\rm{eff}})
\end{equation}
where the effective equation of the state parameter reads
\begin{eqnarray}
w_{\rm{eff}} &=& \frac{1}{1-\ve6\xi(1-6\xi)z^{2}}\Big[ -1 +
\ve(1-6\xi)(1-w_{m})x^{2} \nonumber \\ 
& & + \ve2\xi(1-3w_{m})(x+z)^{2}  
+ (1+w_{m})(1-y^{2}) -
\ve2\xi(1-6\xi)z^{2} - 2\xi\lambda y^{2} z\Big]
\label{eq:weff}
\end{eqnarray}
where $\lambda = -\frac{\sqrt{6}}{\kappa}\frac{1}{U(\phi)}\frac{\ud U(\phi)}
{\ud\phi}$.

The dynamics of the model can be written down as the 4-dimensional autonomous 
dynamical system in variables $x$, $y$, $z$ and $\lambda$ where differentiation 
is with respect to time $\tau$ defined as
$\frac{\ud}{\ud \tau} = \Big[1-\ve6\xi(1-6\xi)z^{2}\Big] \frac{\ud}{\ud \ln{a}}$.\cite{Hrycyna:2010yv} 
However, when the function 
$\Gamma = \frac{\ud^{2}U(\phi)}{\ud \phi^{2}}U(\phi) \left(\frac{\ud U(\phi)}{\ud \phi}\right)^{-2}$ 
is assumed to be the function of $\lambda$ then function $z=z(\lambda)$ can 
be determined from $z(\lambda) =- \int [\lambda^{2}(\Gamma(\lambda)-1)]^{-1} \, \ud\lambda$
and the system can be further reduced to the form of the 
3-dimensional dynamical system 
\begin{eqnarray}
x' & = & -(x-\ve\frac{1}{2}\lambda
y^{2})\Big[1-\ve6\xi(1-6\xi)z(\lambda)^{2}\Big] + 
\frac{3}{2}\left(x+6\xi z(\lambda)\right) 
\bigg[ -\frac{4}{3} - 2\xi \lambda y^{2} z(\lambda) \nonumber \\ & & 
 + \ve(1-6\xi)(1-w_{m})x^{2} 
 +\ve2\xi(1-3w_{m})\left(x+z(\lambda)\right)^{2} + (1+w_{m})(1-y^{2}) \bigg],\\
 y' & = & y\left(2-\frac{1}{2}\lambda x\right)
 \Big[1-\ve6\xi(1-6\xi)z(\lambda)^{2}\Big] 
 +\frac{3}{2} y \bigg[ -\frac{4}{3} - 2\xi \lambda y^{2} z(\lambda) \nonumber \\ & &
 + \ve(1-6\xi)(1-w_{m})x^{2} + 
 \ve2\xi(1-3w_{m})\left(x+z(\lambda)\right)^{2}
 + (1+w_{m})(1-y^{2})\bigg],\\
 \lambda' & = & -\lambda^{2}\left(\Gamma(\lambda)-1\right) x
 \Big[1-\ve6\xi(1-6\xi)z(\lambda)^{2}\Big]
\label{eq:dynsys}
 \end{eqnarray}
if we postulate the exact form of $\Gamma$, for example 
$\Gamma(\lambda) = 1 - \frac{1}{\lambda^{2}}\big(\alpha + \beta\lambda +
\gamma\lambda^{2}\big)$. 
Note that this discussion is not restricted to the specific potential function 
but is generic in the sense that it is valid for any function $\Gamma(\lambda)$ 
for which $z(\lambda)$ exists. In general there are various potential functions 
commonly used in the literature of the subject.\cite{Hrycyna:2010yv}

The qualitative analysis of differential equations allows to study all properties of the solutions without solving the dynamical system. First, we calculate the critical points which correspond mathematically vanishing right-hand sides of the system and physically stationary states. Then we linearize the system around these points. The information about the character of critical point and their stability is contained in eigenvalues of the linearization matrix. If all real parts of eigenvalues are negative (positive) then critical point is stable (unstable). If the all eigenvalues are real of different signs then we have critical point of saddle type organized through the stable and unstable submanifolds.
It is interesting that for our system all important events during the cosmic evolution are represented by these critical points
which traces a generic trajectory representing the evolution of the universe.
Let  us enumerate all these points (see also \cite{Hrycyna:2010yv}). 

1) Finite scale factor singularity; eigenvalues are $l_{1}=6\xi$, $l_{2}=6\xi$, $l_{3}=12\xi$,
and critical point are an unstable node for positive coupling $\xi>0$ and a stable node for negative coupling $\xi<0$.

2a) Fast-roll inflation; $l_1=0$, $l_{2}=12\xi$, $l_{3}=-12\xi$,
and the critical point is non-hyperbolic.

2b) Slow-roll inflation; $l_{1}=l_{2}=l_{3}=0$ and 
the critical point is degenerated.

3) Radiation domination epoch generated by the non-minimal coupling; there are two critical points.
For the phantom scalar field and $\xi>0$ eigenvalues are $l_1=0$, $l_{2}>0$, $l_{3}<0$.
For the canonical scalar field and $\xi>0$ eigenvalues are $l_1=6\xi(1-3w_{m})$,
$l_{2}=12\xi$, $l_{3}=-6\xi$.

4) Matter domination epoch; $l_{1,2}=
-\frac{3}{4}\big[(1-w_{m})\pm [(1-w_{m})^{2}-\frac{16}{3}\xi(1-3w_{m})]^{1/2}\big]$,
$l_{3}=\frac{3}{2}(1+w_{m})$, and the critical points are non-degenerated for $w_{m}\ne-1$ and
$w_{m}\ne\frac{1}{3}$.

5) The present accelerated expansion epoch.

In the most general case without assuming any specific form of the potential
function we are unable to find coordinates of this point. In spite of this we
are able to formulate general conditions for stability of this critical point.
From the Routh-Hurwitz criterion we have that the following conditions should 
be fulfilled to assure stability of this critical point
\begin{equation}
Re{[l_{1,2,3}]}<0 \iff 3\xi\frac{h'(\lambda^{*}_{5})}{z'(\lambda^{*}_{5})}
(y^{*}_{5})^{2} >0 \quad \textrm{where} \quad 
h(\lambda)= \lambda z(\lambda)^{2} +4 z(\lambda) -\frac{\lambda}{\ve6\xi}.
\label{eq:stabcon2}
\end{equation}

\begin{figure}
\begin{center}
\includegraphics[scale=0.72]{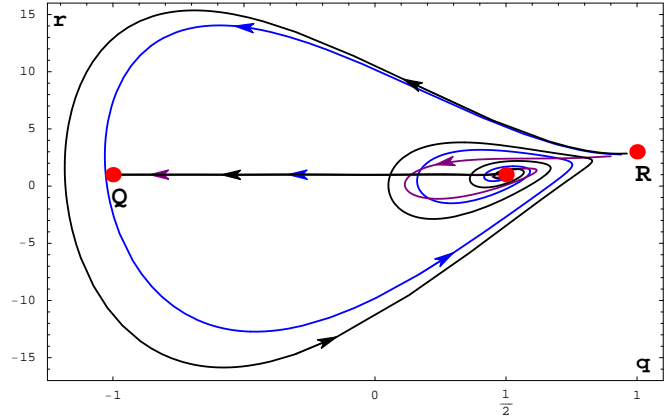}
\end{center}
\caption{The state-finder parameter diagnostic for the twister quintessence
scenario and $z(\lambda)=\frac{\lambda}{\alpha}$.
The trajectories (black $\xi=6$, $\alpha=-6$, blue $\xi=4$, $\alpha=-4$ and
purple $\xi=2$, $\alpha=-2$) represent a twister type solution which
interpolates between $R$ -- the radiation dominated universe (a saddle type critical point), the matter dominated universe (an unstable focus critical point) and $Q$ -- the accelerating universe (a stable critical point).}
\label{fig:1}
\end{figure}

To emphasize the acceleration epoch in the twister solution, the state-finder 
parameters are calculated: $q=-\ddot{a}/aH^{2}$ is the
deceleration parameter and $r=\dddot{a}/aH^3$ (see Fig.~\ref{fig:1}).

\section{Conclusions}

In this paper we pointed out the presence of the new interesting solution for 
the non-minimally coupled scalar field cosmology which we called the twister
solution (because of the shape of the corresponding trajectory in the phase
space). This type of the solution is very interesting because in the phase space
it represents the 3-dimensional trajectory which interpolates different stages of 
evolution of the universe, namely, the radiation dominated, dust filled and
accelerating universe. We are able to find linearized solutions around all these 
intermediate phases, and hence, parameterizations for $w_{\rm{eff}}(a)$ in 
different epochs of the universe history. It is interesting that the presented 
structure of the phase space is allowed only for non-zero value of the coupling 
constant, therefore it is a specific feature of the non-minimally coupled scalar 
field cosmology. Our general conclusion is that in the description of dynamical
complexity the cosmic evolution including both barotropic matter and
non-minimally scalar field leads to a new richer dynamics.

\end{document}